\documentclass[aps,pra,twocolumn,superscriptaddress,showpacs]{revtex4}

\usepackage{graphicx}
\usepackage{epstopdf}
\usepackage{tabularx}
\usepackage{amssymb}
\usepackage{amsmath}
\usepackage{bm}
\usepackage{graphpap}
\usepackage{multirow}
\usepackage{notes2bib}
\usepackage{color}
\usepackage[version=3]{mhchem} % Formula subscripts using \ce{}

\begin{document}
%%%%%%%%%%%%%%%%%%%%%%%%%%%%%%%%%%%%%%%%%%%%%%%%%%%%%
\title{Atomic signatures of local environment from core spectroscopy in $\beta$-\ce{Ga2O3}}
%%%%%%%%%%%%%%%%%%%%%%%%%%%%%%%%%%%%%%%%%%%%%%%%%%%%%%
\author{Caterina \surname{Cocchi}}
\email{caterina.cocchi@physik.hu-berlin.de}
\affiliation{Physics Department, Humboldt-Universit\"at zu Berlin, Berlin, Germany}
\affiliation{IRIS Adlershof, Humboldt-Universit\"at zu Berlin, Berlin, Germany}
\affiliation{European Theoretical Spectroscopic Facility (ETSF)}
\author{Hannes \surname{Zschiesche}}
\affiliation{Physics Department, Humboldt-Universit\"at zu Berlin, Berlin, Germany}
\author{Dmitrii \surname{Nabok}}
\affiliation{Physics Department, Humboldt-Universit\"at zu Berlin, Berlin, Germany}
\affiliation{IRIS Adlershof, Humboldt-Universit\"at zu Berlin, Berlin, Germany}
\affiliation{European Theoretical Spectroscopic Facility (ETSF)}
\author{Anna \surname{Mogilatenko}}
\affiliation{Physics Department, Humboldt-Universit\"at zu Berlin, Berlin, Germany}
\affiliation{Ferdinand-Braun-Insititut, Leibniz-Institut f\"ur H\"ochstfrequenztechnik, Berlin, Germany}
\author{Martin \surname{Albrecht}}
\affiliation{Leibniz Institute for Crystal Growth, Berlin, Germany}
\author{Zbigniew \surname{Galazka}} 
\affiliation{Leibniz Institute for Crystal Growth, Berlin, Germany}
\author{Holm \surname{Kirmse}}
\affiliation{Physics Department, Humboldt-Universit\"at zu Berlin, Berlin, Germany}
\author{Claudia \surname{Draxl}}
\affiliation{Physics Department, Humboldt-Universit\"at zu Berlin, Berlin, Germany}
\affiliation{IRIS Adlershof, Humboldt-Universit\"at zu Berlin, Berlin, Germany}
\affiliation{European Theoretical Spectroscopic Facility (ETSF)}
\author{Christoph T. \surname{Koch}}
\affiliation{Physics Department, Humboldt-Universit\"at zu Berlin, Berlin, Germany}

%%%%%%%%%%%%%%%%%%%%%%%%%%%%%%%%%%%%%%%%%%%%%%%%%%%%%%
\date{\today}
\pacs{71.35.-y, 73.20.Mf, 79.20.Uv}
\begin{abstract}
We present a joint theoretical and experimental study on core-level excitations from the oxygen $K$-edge of $\beta$-\ce{Ga2O3}.
A detailed analysis of the electronic structure reveals the importance of O-Ga hybridization effects in the conduction region.
The  spectrum from O 1$s$ core electrons is dominated by excitonic effects, which overall red-shift the absorption onset by 0.5 eV, and significantly redistribute the intensity to lower energies.
Analysis of the spectra obtained within many-body perturbation theory reveals atomic
fingerprints of the inequivalent O atoms.
From the comparison of energy-loss near-edge fine-structure (ELNES) spectra computed with respect to different crystal planes, with measurements recorded under the corresponding diffraction conditions, we show how the spectral contributions of specific O atoms can be enhanced while quenching others.
These results suggest ELNES, combined with \textit{ab initio} many-body theory, as a very powerful technique to characterize complex systems, with sensitivity to individual atomic species and to their local environment.
\end{abstract}
\maketitle
%%%%%%%%%%%%%%%%%%%%%%%%%%%%%%%%%%%%%%%%%%%%%%%%%%%%%
% INTRODUCTION
%%%%%%%%%%%%%%%%%%%%%%%%%%%%%%%%%%%%%%%%%%%%%%%%%%%%%
\section{Introduction}

Transparent conductive oxides (TCOs) are very promising materials for opto-electronics, due to their wide band gap \cite{grundmann_TCO-materials-devices}, transparency to visible light, and good n-type conductivity \cite{robertson_TCO-electric-structure_2008,hajnal_oxygen-vacancy-Ga2O3}.
Recent advances in the synthesis of group-III sesquioxides single crystals \cite{aida+08jjap,gala+10crt,gala+13jcg,galazka_Ga2O3_2014} have drawn particular attention to these compounds.
$\beta$-\ce{Ga2O3} is one of the most interesting ones, due to its unique electronic and optical properties. 
With its band gap of about $4.8$ eV \cite{gala+10crt} and its remarkably high carrier mobility \cite{thom+14apl}, it is considered an excellent candidate for the new generation of opto-electronic devices.
These properties can be further engineered by forming ternary compounds, upon inclusion of \textit{e.g.}, indium atoms \cite{zhang_wide-bandgap-engineering-GaIn2O3-films_2014}.
In order to achieve full control in tailoring these materials with increasing level of complexity, it is necessary to rigorously characterize and fully understand the features of the pristine systems.

Electron energy loss spectroscopy (EELS) is a powerful technique in this respect.
In the transmission electron microscope (TEM) core electrons can be excited when probe electrons are inelastically scattered in their proximity, giving rise to energy losses at the ionization edge.
In this way, through the so-called energy loss near-edge fine structure (ELNES), fingerprints of chemical environment and electronic structure of excited atoms can be detected.
Scanning transmission electron microscopy (STEM) EELS allows for the investigation of the electronic structure of selected crystal regions, namely columns of atoms, by scanning over the sample \cite{kura+09jm}. 
In comparison to STEM, illuminating the specimen by a wide beam in the TEM mode offers a lower risk to manipulate or damage the sample by a too high dose of impinging electrons,  and does not require an aberration corrector for the condenser optics of the microscope.
To selectively probe specific crystal regions or planes in the TEM mode measurements can be performed in diffraction mode, by means of the energy loss by channeled electrons (ELCE) technique \cite{tftoe_ssELNES_1982}. 
Aligning the crystal in a systematic row condition with respect to the incident beam establishes standing waves with maxima of the electron intensity at a well-defined set of crystal planes. In this case, the ELNES signal becomes highly sensitive to energy losses in different regions of the unit cell upon small variations of the diffraction conditions \cite{tatsumi_ssELNES_2009}, or of the momentum transfer of the detected signal \cite{hetaba_ssEFSrutile_2014}.
ELNES fingerprints of O 1$s$ electrons are of special importance, since they differ between pure $\beta$-\ce{Ga2O3} crystals and other group-III sesquioxides, such as \ce{In2O3}, and their ternary compounds, such as $\text{Ga}_{2(1 - x)}\text{In}_{2x}\text{O}_3$. 

Polarization-dependent core spectroscopy represents another technique to investigate local electronic structures beyond the use of electrons.
While in x-ray absorption spectroscopy the polarization of the illumination can be utilized as an additional degree of freedom for obtaining more detailed structural insight \cite{sche+95prb,stoh99jmmm,seki+13jesrp,wang+14jesrp}, this is not possible with high-energy electrons. 
However, the strong elastic scattering cross section of electrons in the matter allows for orientation-sensitive electron channeling, which is the basis for the ELCE approach, to be employed as an additional channel of structural information.

A thorough interpretation and understanding of ELNES and ELCE features requires insight from theory.
The complexity of the excited-state spectra of TCOs \cite{yama+11pssc,mich-schm12iopcs,stur+15aplm} calls for an accurate first-principles description, that explicitly treats many-body effects. 
In this paper, we present a combined theoretical and experimental work investigating 
ELNES fingerprints of the local atomic environment in $\beta$-\ce{Ga2O3}.
A detailed characterization of the electronic structure of the system is the prerequisite to analyze the core-excitation spectrum from the O $K$-edge.
We demonstrate the relevance of excitonic effects for correctly reproducing the experimental results, and analyze the character of the electron-hole ($e$-$h$) pairs.
Finally, we compare measured ELNES spectra, which are sensitive to the selected diffraction conditions, with the theoretical results, shedding light on the fingerprints of individual oxygen atoms.

This paper is organized as follows.
In Section \ref{sec:sys+meth} we introduce the crystal structure of $\beta$-\ce{Ga2O3} (Section \ref{sec:system}), as well as the theoretical (Section \ref{section:theory}), computational (Section \ref{section:comput}), and experimental methods (Section \ref{sec:exp}) employed in this study.
In Section \ref{sec:electronic}, we present the electronic properties of $\beta$-\ce{Ga2O3}, and in Section \ref{sec:xas} we analyze its core-excitation spectrum from the O $K$-edge, and discuss the character of the lowest-energy $e$-$h$ pair.
Finally, in Section \ref{sec:ELNES},  we analyze the features of the theoretical ELNES in comparison with the corresponding experimental spectra, highlighting the contributions of each inequivalent oxygen atom, depending on the specific diffraction conditions.

%FIGURE 1
\begin{figure}
\center
\includegraphics[scale=0.3]{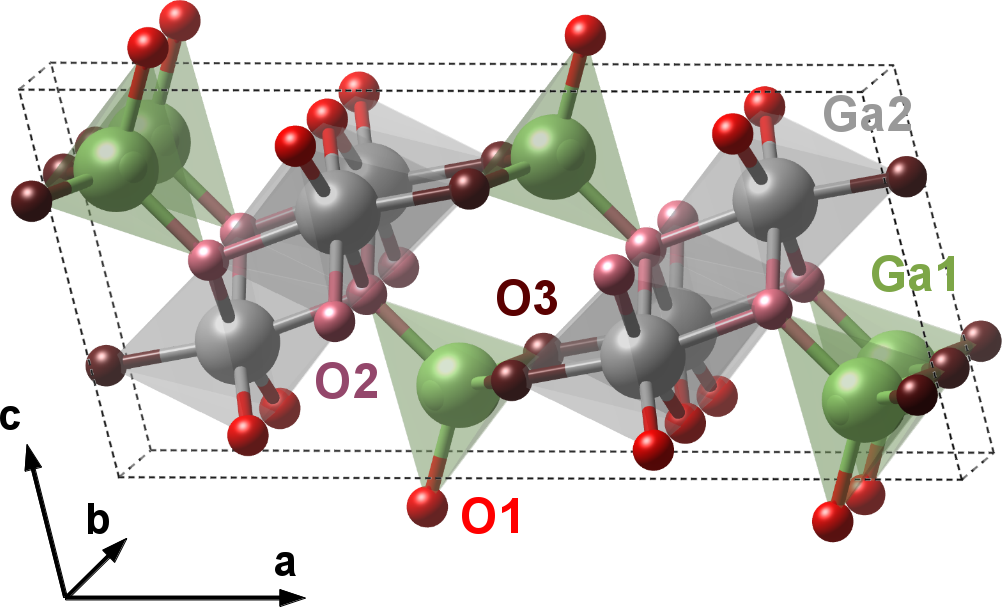}
\caption{(Color online) Unit cell of $\beta$-\ce{Ga2O3}. Tetrahedrally coordinated gallium atoms (Ga1) are depicted in green and octahedrally coordinated ones (Ga2) in grey. Inequivalent O atoms are indicated in red (O1), pink (O2), and maroon (O3). 
\label{fig:Ga2O3-structure}
}
\end{figure} 
%

%%%%%%%%%%%%%%%%%%%%%%%%%%%%%%%%%%%%%%%%%%%%%%%%%%%%%
%  SYSTEMS & METHODS
%%%%%%%%%%%%%%%%%%%%%%%%%%%%%%%%%%%%%%%%%%%%%%%%%%%%% 
\section{System and Methods}
\label{sec:sys+meth}
%%%%%%%%%%%%%%%%%%%%%%%%%%
\subsection{Crystal structure of $\beta$-\ce{Ga2O3}}
\label{sec:system}

%
%TABLE 1
\begin{table}
\caption{Coordination of each inequivalent oxygen atoms in the unit cell of $\beta$-\ce{Ga2O3}, and number of bonds with differently coordinated Ga atoms.
\label{tab:ga2o3-oxygenbonds}
}
\center
\begin{ruledtabular}
\begin{tabularx}{\textwidth}{lccc}
        Atom & coordination & tetrahedral Ga1 & octahedral Ga2 \\
        \hline
        O1 & & 1 & 2 \\
        O2 & octahedral & 1 & 3 \\
        O3 & tetrahedral & 2 & 1 \\
\end{tabularx}
\end{ruledtabular}
\end{table}

In the $\beta$-phase, \ce{Ga2O3} has a monoclinic crystal structure with lattice parameters $a$=12.23 \AA{}, $b$=3.04 \AA{}, $c$=5.80 \AA{}, and a monoclinic angle $\beta$=103.7$^{\circ}$ \cite{gell60jcp,ahman_Ga2O3-lattice_1996}.
Its unit cell contains 20 atoms, including two inequivalent Ga atoms and three inequivalent O atoms, with different chemical coordination (see Fig. \ref{fig:Ga2O3-structure} and Tab. \ref{tab:ga2o3-oxygenbonds}).
Tetrahedrally and octahedrally coordinated gallium atoms are present, labeled as Ga1 and Ga2, respectively.
Inequivalent O atoms are classified according to their bonding with the neighboring Ga ions, and are associated with different colors in Fig. \ref{fig:Ga2O3-structure}.
O1 (red) share two bonds with Ga2 and one bond with Ga1, while O2 (pink) are predominantly connected to octahedral Ga2 (3 bonds) and have only one bond to Ga1.
O3 (maroon) exhibit two bonds to Ga1 and only one to Ga2.

% FIGURE 2
\begin{figure}
\center
\includegraphics[scale=0.3]{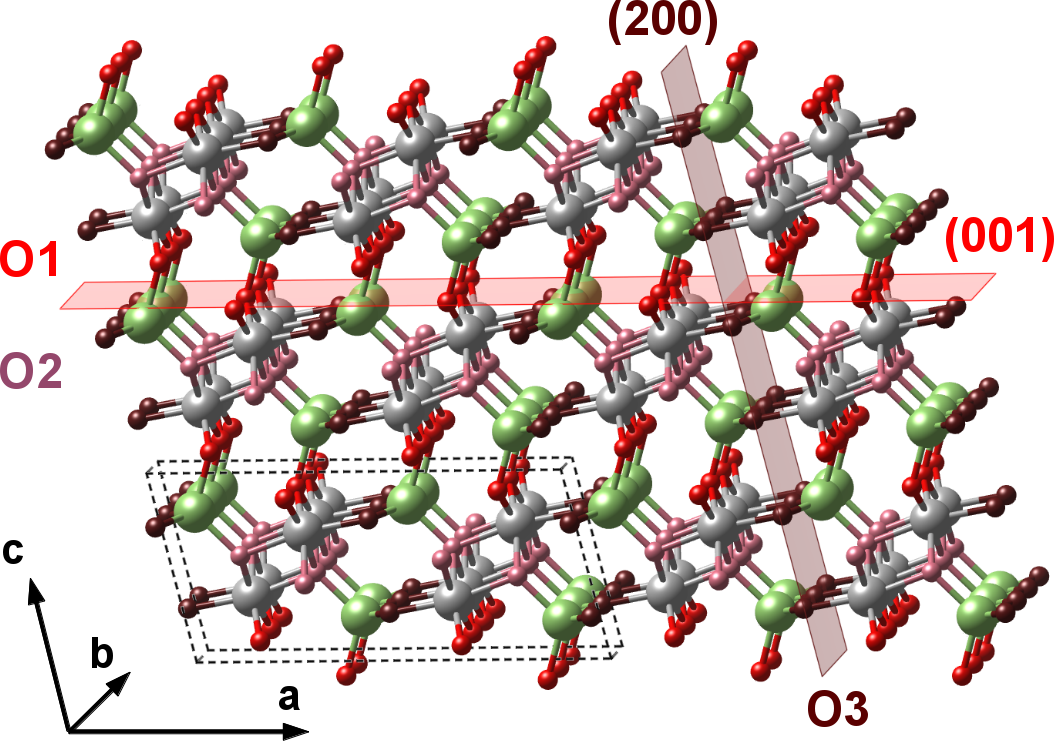}
\caption{(Color online) $\beta$-\ce{Ga2O3} crystal in [010] direction. (200) and (001) crystal planes are highlighted. The unit cell of $\beta$-\ce{Ga2O3} is indicated by dashed lines.
\label{fig:Ga2O3-planes}
}
\end{figure}

In Fig. \ref{fig:Ga2O3-planes} we highlight two crystal planes that contribute to the diffraction pattern in the [010] zone axis, as considered in our experimental setup.
In the (200) crystal plane there is an alternation of planes, which are packed with either O3 or O2 and O1. 
\textit{Pure} planes containing only O3 are characterized by lower atomic density, compared to the so-called \textit{mixed} planes, containing O2, O1, and gallium atoms at the same time.
On the other hand, (001) crystal planes are alternately packed with only one inequivalent oxygen atom. 
However, in this case, \textit{pure} planes contain either O2 or O1, while \textit{mixed} ones include O3 and gallium atoms.
The same occurs in the (201) plane (not shown), which additionally contributes to the diffraction pattern along the [010] crystallographic direction.

%%%%%%%%%%%%%%%%%%%%%%%%%%%%%
\subsection{Theoretical background}
\label{section:theory}

Core-level excitations are computed from first principles by solving the Bethe-Salpeter equation (BSE), an effective equation of motion for the electron-hole two-particle Green's function \cite{hank-sham80prb,stri88rnc}.
Considering that transitions occur from core ($c$) to unoccupied ($u$) states, the BSE, in its matrix form, can be written as:
\begin{equation}
\sum_{c'u'\mathbf{k'}} \hat{H}^{BSE}_{cu\mathbf{k},c'u'\mathbf{k'}} A^{\lambda}_{c'u'\mathbf{k'}} = E^{\lambda} A^{\lambda}_{cu\mathbf{k}} .
\label{eq:BSE}
\end{equation}
Here we are considering only excitations from the O $K$ edge, thus O 1$s$ are the only initial states.
The BSE Hamiltonian in Eq. \ref{eq:BSE} is given by the sum of three terms: 
\begin{equation}
\hat{H}^{BSE} = \hat{H}^{diag} + 2 \gamma_x \hat{H}^x + \gamma_c \hat{H}^{dir}.
\label{eq:H_BSE}
\end{equation}
The \textit{diagonal} term $\hat{H}^{diag}$ accounts for single-particle transitions, while the \textit{exchange} ($\hat{H}^x$) and \textit{direct} ($\hat{H}^{dir}$) terms in Eq. \ref{eq:H_BSE} incorporate the repulsive bare and the attractive statically screened Coulomb interaction, respectively.
The coefficients $\gamma_x$ and $\gamma_c$ in Eq. \ref{eq:H_BSE} enable to select the \textit{spin-singlet} ($\gamma_x$ = $\gamma_c$ = 1) and \textit{spin-triplet} ($\gamma_x$ = 0, $\gamma_c$ = 1) channels.
In the latter case, the exchange interaction is not present.
In Eq. \ref{eq:BSE}, the eigenvalues $E^{\lambda}$ represent excitation energies.
The exciton binding energy ($E_b$) is assumed with respect to the absorption onset that is obtained from the independent-particle approximation (IPA), through the diagonalization of $\hat{H}^{diag}$.
The eigenvectors $A^{\lambda}$ carry information about the character and the composition of the excitons.
In fact, they explicitly appear as coefficients in the expression of the two-particle wave-function:
\begin{equation}
\Psi^{\lambda}(\mathbf{r}_h,\mathbf{r}_e)= \sum_{cu\mathbf{k}} A^{\lambda}_{cu\mathbf{k}} \phi_{uk}(\mathbf{r}_e)\phi^*_{ck}(\mathbf{r}_h) ,
\label{eq:exciton-wf}
\end{equation}
where $\phi_{ck}(\mathbf{r}_h)$ and $\phi_{uk}(\mathbf{r}_e)$ are the Kohn-Sham core and conduction states, respectively.
Moreover, through the oscillator strength, given by the square modulus of
\begin{equation}
\mathbf{t}_{\lambda}= \sum_{cu\mathbf{k}} A^{\lambda}_{cu\mathbf{k}} \dfrac{\langle c\mathbf{k}|\widehat{\mathbf{p}}|u\mathbf{k}\rangle}{\epsilon_{u\mathbf{k}} - \epsilon_{c\mathbf{k}}} ,
\label{eq:t}
\end{equation}
$A^{\lambda}$ enter the expression of the imaginary part of the macroscopic dielectric function ($\varepsilon_M$):
\begin{equation}
\mathrm{Im}\varepsilon_M = \dfrac{8\pi^2}{\Omega} \sum_{\lambda} |\mathbf{t}_{\lambda}|^2 \delta(\omega - E_{\lambda}) ,
\label{eq:ImeM}
\end{equation}
which yields the absorption of the system.
The loss function, describing EELS, is proportional to the inverse of this quantity: $L(\omega)=-\mathrm{Im}(1/\varepsilon_M)$.
Details on the BSE in the all-electron framework of the linearized augmented planewave (LAPW) method can be found in Refs. \cite{pusc-ambr02prb,Sagmeister2009}.

An alternative approach to compute core excitations is given by the so-called supercell core-hole approximation, where the excitation is mimicked by eliminating the core electron and by adding it to the conduction band \cite{soin-shir01prb,dusc+01um}.
Within this method, the $e$-$h$ Coulomb interaction is taken into account self-consistently at the density-functional theory (DFT) level, yielding good results for deep core excitations, where correlation effects are not prominent \cite{rehr+05ps}.
On the other hand, this method is known to perform worse for excitations from shallow core or semicore states, where $e$-$h$ correlation becomes relevant \cite{rehr+05ps,olov+09jpcm,olov+11prb,olov+13jpcm}.
In the core-hole approximation, the exciton binding strength relies on the adopted approximation for the exchange-correlation functional \cite{olov+09prb}.
Moreover, this method requires the construction of a supercell to avoid spurious interactions between neighboring core-holes.
Hence, for materials with large unit cells, such as $\beta$-\ce{Ga2O3}, such calculations become as costly as BSE.
Consequently, here we adopt the BSE approach, which is the state of the art in this field, and which also provides us with exciton binding energies and the corresponding $e$-$h$ wave-functions.

%%%%%%%%%%%%%%%%%%%%%%%%%%%%%
\subsection{Computational details}
\label{section:comput}

All calculations are performed in the framework of DFT and many-body perturbation theory, applying the all-electron full-potential LAPW method, as implemented in the \texttt{exciting} code \cite{gula+14jpcm}.
Starting from the experimental crystal structure of $\beta$-\ce{Ga2O3} \cite{gell60jcp}, atomic positions are optimized in order to obtain forces smaller than 50 meV/\AA{}.
The muffin-tin radii of Ga and O are set to 1.85 bohr and 1.60 bohr, respectively. 
A plane-wave cutoff $R_{\textrm{MT}}G_{\textrm{max}} = 8$ is applied to the basis set, and the Brillouin zone is sampled by a $2\times8\times4$ \textbf{k}-point grid. 
The local-density approximation (LDA, Perdew-Wang functional \cite{perd-wang92prb}) is used for the exchange-correlation potential.
The BSE is solved on a shifted $2\times8\times4$ \textbf{k}/\textbf{q}-point grid, which ensures convergence up to 150 meV of the binding energy of the first exciton.
We include transitions from O 1$s$ states to 48 unoccupied states spanning an energy range of approximately 16 eV above the absorption onset.
The screened Coulomb interaction between the electron and the hole is computed within the random-phase approximation, by including the initial O 1$s$ states, all valence bands, and 100 empty bands.
Local-field effects are included considering 32 $|\mathbf{G}+\mathbf{q}|$ vectors.
A scissors operator of 25 eV is applied to the BSE spectra to align the energy of the first and most intense peak to the experimental counterpart.
The same shift is also applied to the IPA spectra.
For visualization of the crystal structure we make use of the VESTA software \cite{momm-izum11jacr}.

To simulate ELNES spectra of $\beta$-\ce{Ga2O3} with respect to different crystal planes, we solve Maxwell's equations starting from the four independent dielectric tensor components computed from BSE, in account of the anisotropy of the monoclinic crystal structure.
To do so, we employ \texttt{LayerOptics} \cite{vorw+16cpc}, a code implementing the $4 \times 4$ matrix-formalism for the solution of Maxwell's equations in anisotropic layered materials \cite{yeh80ss,pusc-ambr06aem}.
Details on the formalism and on the computational procedure are reported in Ref. \cite{vorw+16cpc}.
To reproduce different diffraction conditions, we vary the orientation of the sample with respect to the considered crystal planes.
This corresponds to an effective rotation of the four components of the dielectric tensor by Euler angles $\alpha=19.97^{\circ}$, $\beta=94.95^{\circ}$, and $\gamma=257.6^{\circ}$ for the (001) crystal plane, and by $\beta=131^{\circ}$ for the (200) plane.
The impinging electron beam is normal to the surface of the sample, considered with a thickness of 50 nm.

%%%%%%%%%%%%%%%%%%%%%%%%%%%%%
\subsection{Experimental methods}
\label{sec:exp}
%%%%%%%%%%%%%%%%%%%%%%%%%%%%%%%%%%%%%%%%%%%%%%%%%%%%%%%%%%%%%%%%%%%%%%%%%%%%%%%%%%%%%%%%%%%%%%%
Bulk $\beta$-\ce{Ga2O3} single crystals are grown from the melt by the Czochralski method \cite{gala+10crt,galazka_Ga2O3_2014}. 
The crystal shows n-type conductivity with free electron concentration of 5 $\times$10$^{17}$ cm$^{-3}$ and free electron mobility of 110 cm$^2$V$^{-1}$s$^{-1}$, as determined by Hall effect measurements.
Thin lamella normal to the [010] zone axis is prepared for the TEM investigation, by plan parallel polishing and Ar$^+$ ion milling in a liquid nitrogen cooled Gatan PIPS \cite{galazka_Ga2O3_2014}. 
TEM experiments are performed using a JEOL JEM 2200FS. 
The microscope is equipped with an in-column energy filter and a 1k slow-scan CCD camera (Gatan). 
The operating acceleration voltage is 200 kV. 
The specimen is tilted slightly off the [010] zone axis in order to achieve systematic row conditions strongly exciting either the (200) or the (001) diffraction spot. 
This ensures strong interaction with the selected crystal planes.
The detector is positioned on the direct beam to get results comparable with theory. 
The illumination convergence semi-angle is set to 1.6 mrad, and the collection semi-angle to 2.2 mrad, at a camera length of 150 cm. 
This fairly parallel illumination and small detection angle are chosen to make experimental and theoretical data comparable without extra computational effort.    
The dispersion is chosen as $200\ \text{eV}\mu\text{m}^{-1}$. 
The acquisition time of one frame is $30\ \text{s}$. 
The energy resolution of the spectra is determined to be 1.2 eV by measuring the full width at half maximum of the zero loss peak. Multiple spectra are collected and averaged after aligning them for potential energy shifts.
The core-loss spectrum is corrected for multiple scattering by deconvolution with a spectrum recorded in the low-loss regime.
The background signal subtracted from the experimental data is extrapolated according to a power law fitted to a region in the spectrum about 20 eV below the oxygen $K$-edge (routine provided by the Gatan Digital Micrograph software).

%%%%%%%%%%%%%%%%%%%%%%%%%%%%%%%%%%%%%%%%%%%%%%%%%%%%%%%%%%%%%%%%%%%%%%%%%%%%%%%%%%%%%%%%%%%%%%%

%%%%%%%%%%%%%%%%%%%%%%%%%%%%%%%%%%%%%%%%%%%%%%%%%%%%%
%  RESULTS and DISCUSSION
%%%%%%%%%%%%%%%%%%%%%%%%%%%%%%%%%%%%%%%%%%%%%%%%%%%%
\section{Results and Discussion}

\subsection{Electronic properties}
\label{sec:electronic}

% FIGURE 2
\begin{figure}
\centering
\includegraphics[width=.45\textwidth]{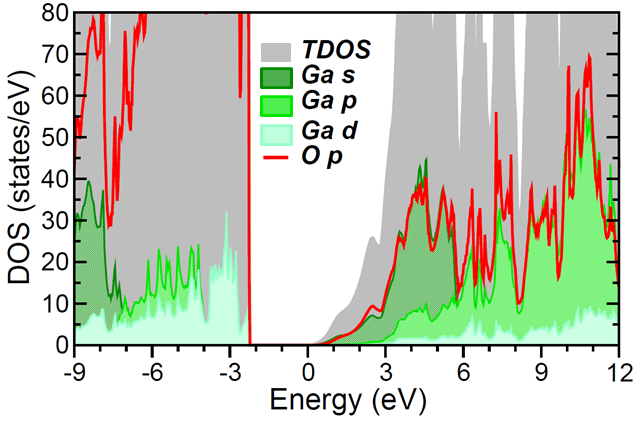}%
\caption{(Color online) Density of states (DOS) of $\beta$-\ce{Ga2O3}, including the total contribution (TDOS) as well as the projections on Ga $s$, $p$, $d$, and O $p$ states. The Fermi energy is set to 0 eV.}
\label{fig:bands}
\end{figure}

The density of states (DOS) of $\beta$-\ce{Ga2O3} is shown in Fig. \ref{fig:bands}. 
This result is obtained at the DFT level.
Hence, the band gap of about 2.3 eV is naturally underestimated with respect to the experimental value \cite{moha+10apl}.
To reproduce the correct absorption onset, a scissors operator of 25 eV is applied to the excited-state spectra (see also Section \ref{section:comput}).
From Fig. \ref{fig:bands}, we can extract relevant information about the character of the electronic states, which is crucial in view of understanding the nature of core-level excitations. 
The valence region is characterized by a manifold of bands with predominant O-$p$ character, in agreement with previous \textit{ab initio} calculations, performed with LDA \cite{liti+09jac}, as well as with hybrid functionals \cite{he+06prb}.
In the same energy interval we find also contributions from Ga atoms, with $p$ and especially $d$ states close to the gap.
The bottom of the conduction region is dominated by hybridized states, with O-$p$ and Ga-$s$ character.
At higher energies, at about 2 eV above the Fermi energy ($E_F$), the unoccupied bands are again strongly hybridized, involving in this case O-$p$ and Ga-$p$ states.
Ga-$d$ bands, which dominate the top of the valence band, have little weight in the considered energy range above $E_F$.
In the core region (not shown), 1$s$ electrons of the three inequivalent O atoms are separated by approximately 0.2 eV, with O1 being the shallowest and O2 the deepest. 
%%%%%%%%%%%%%%%%%%%%%%%%%%%%%%%%%%%
\subsection{Core-level absorption spectrum from the O $K$-edge}
\label{sec:xas}

% FIGURE 3
\begin{figure}
\centering
\includegraphics[width=.48\textwidth]{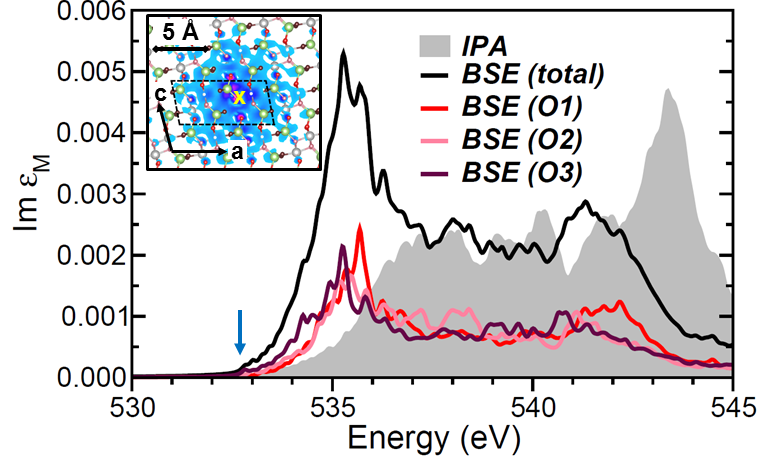}%
\caption{(Color online)  Core-level absorption spectrum from the O $K$-edge of $\beta$-\ce{Ga2O3}, given by the average of the diagonal components of Im$\varepsilon_M$. BSE results are indicated by solid lines, including contributions from the inequivalent O atoms, as well as from their sum. The independent-particle (IPA) spectrum is shown for comparison (shaded area). A Lorentzian broadening of 0.1 eV is applied to all spectra. Inset: Two-dimensional projection on the (010) plane of the electron distribution of the lowest-energy exciton from the 1$s$ core state of O2, indicated by the arrow. The position of the hole is marked by a cross.}
\label{fig:XAS}
\end{figure}
In Fig. \ref{fig:XAS}, the core-level absorption spectrum from the O $K$-edge is shown, taken as an average of the diagonal components of Im$\varepsilon_M$.
In the BSE spectra (solid lines), we indicate separately the contributions from the inequivalent O atoms, according to the color code already adopted for the crystal structure in Figs. \ref{fig:Ga2O3-structure} and \ref{fig:Ga2O3-planes}: O1 (red), O2 (pink), and O3 (maroon).
Their sum is depicted in black.
For analysis purposes, we include also the IPA spectrum (grey shaded area).
At the onset three bound excitons appear.
The first one ($\lambda$=1) has a binding energy $E_b \simeq$ 0.5 eV.
This result is comparable with binding energies of core-excitations from the Li and Be $K$-edge in solids, such as \ce{LiS2} \cite{olov+09jpcm} and \ce{BeS} \cite{olov+13jpcm}.
Conversely, near-edge core-excitons in organic systems exhibit larger binding energies, in the order of a few eV \cite{cocc-drax15prb}.
The lowest-energy excitation in $\beta$-\ce{Ga2O3} targets the conduction-band minimum.
In the inset, the exciton wave-function (Eq. \ref{eq:exciton-wf}), corresponding to the transition from O2 1$s$, is depicted.
Specifically, for a fixed position of the hole, marked by a cross, we show the extension of the electron distribution of $|\Psi^{\lambda=1}(\bar{\mathbf{r}_h}, \mathbf{r}_e)|^2$, localized within a few next-nearest-neighbor sites.
The near-edge region, starting at about 535 eV, is characterized by an intense excitonic peak, stemming from contributions of all inequivalent oxygen atoms.
The maxima of the BSE spectra, considering separately transitions from O1, O2, and O3, span a range of approximately 1 eV. 
At the onset, the spectrum from O3 exhibits relatively intense features, with a peak at $\sim$535.5 eV, a shoulder at 535 eV, and a near-edge excitation at 534 eV.
The absorption from O1 has its maximum at about 536 eV.
The spectrum from O2 shows weaker peaks, compared to the contributions from the other O atoms.
Its maximum is located between the most intense peaks obtained from O1 and O3 $K$-edges.
At higher energies (537 -- 540 eV), the overall intensity decreases.
Between 541 and 543 eV, the spectrum features another peak, with approximately half the intensity of the first one.
The spectrum from O1 shows a broader excitation band in this region, while O2, and especially O3, exhibit a sharper peak at about 541 eV.
Our BSE spectrum, with the first peak at about 535 eV being more intense than the second one, is in agreement with recent x-ray absorption measurements \cite{mich-schm12iopcs}.

The comparison with the IPA spectrum clarifies the excitonic nature of the main peaks.
The inclusion of the $e$-$h$ interaction induces a red-shift of the absorption onset by 0.5 eV, corresponding to the binding energy of the first exciton. Moreover, the first peak is red-shifted by about 2 eV, and the spectral weight is significantly redistributed to lower energy.
In fact, in the IPA picture, the lower-energy peak is less intense than the higher-energy one.
This spectral shape reflects the O-$p$ contributions to the unoccupied DOS (Fig. \ref{fig:bands}), which shows strong hybridization with Ga-$s$ states in the lowest-energy part, and with Ga-$p$ ones about 6 eV above $E_F$.
Thus, the $e$-$h$ interaction tends to increase the spectral weight of the excitations targeting O-$p$ states hybridized with Ga-$s$ ones, while quenching the oscillator strength of transitions to more localized Ga-$p$ bands.
The inclusion of the $e$-$h$ interaction does not modify the character of the excitations, despite such strong redistribution of intensity.
Excitons associated with the first peak still involve unoccupied O-$p$ states hybridized with Ga-$s$.
Analogously, the second peak corresponds to transitions to hybridized O-$p$/Ga-$p$ bands.
Excitations related to both peaks are rather delocalized, reflecting the character of the final unoccupied states.

%%%%%%%%%%%%%%%%%%%%%%%%%%%%%%%%%%%
\subsection{ELNES spectra}
\label{sec:ELNES}

% FIGURE 4
\begin{figure}
\centering
\includegraphics[width=.48\textwidth]{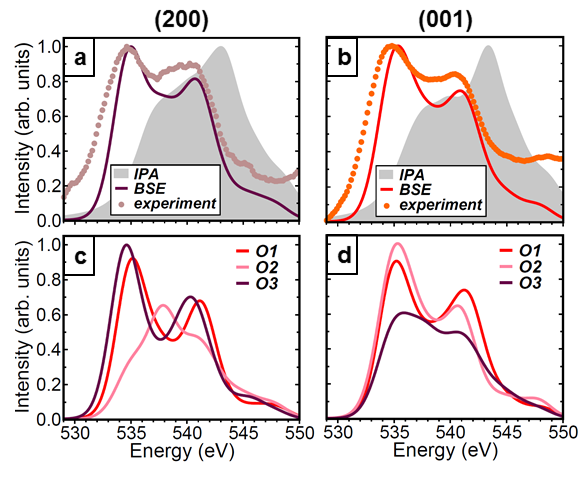}%
\caption{(Color online) ELNES spectra of $\beta$-\ce{Ga2O3} along the (200) and (001) crystal planes, showing (a,b) the total theoretical spectra (BSE, solid lines and IPA, shaded area), as well as the experimental ones (dots); (c,d) Oxygen-resolved contributions. The computed spectra are convolved with a Lorentzian broadening of 1.2 eV. The intensity of all the spectra is normalized to the height of their strongest peak.
}
\label{fig:ELNES}
\end{figure}

With the knowledge of the core-level absorption spectra from the inequivalent O atoms, we can finally analyze the ELNES spectra.
In Fig. \ref{fig:ELNES}a-b we report our theoretical and experimental results obtained for the sample oriented along the (200) and (001) crystal planes.
The energy range is chosen to show a direct and meaningful comparison between measured and computed spectra \bibnote{The pre-edge shoulder appearing in the spectrum in Fig. \ref{fig:ELNES}a is an artifact in the experimental data, potentially due to afterglow of the scintillator of the charge-coupled device (CCD) camera.}. 
The main features discussed in Fig. \ref{fig:XAS} for the average core-level spectrum from the O $K$-edge appear also here, irrespective of the crystallographic orientation.
The strong anisotropy of $\beta$-\ce{Ga2O3} does not overall affect excitation energies and intensity.
The two main peaks are pronounced in both cases, with the higher-energy one having 80$\%$ of the intensity of the lower-energy one.
The correct spectral shape is not reproduced by the IPA result, which is not only quantitatively, but also qualitatively wrong.
The good agreement between theory and experiment confirms the importance of including the $e$-$h$ interaction to describe absorption spectra of TCOs.

The situation is quite different when considering separately the contributions from inequivalent oxygen atoms (Fig. \ref{fig:ELNES}c-d).
Due to their different coordination in the unit cell of $\beta$-\ce{Ga2O3}, we expect the spectral weight to vary significantly from one crystal plane to the other, depending on the initial core state.
Indeed, while the spectrum from O1 is the same in both considered orientations, transitions from O2 and O3 are remarkably sensitive to the diffraction condition.
Along the (200) plane the oscillator strength in the spectrum computed from O2 is distributed differently compared to the other oxygen contributions.
A maximum appears at $\sim$539 eV, where the ELNES of O1 and O3 show a local minimum.
This feature is also reflected in the theoretical and experimental spectra shown in Fig. \ref{fig:ELNES}a.
In  both cases the intensity between the first and second peak does not decrease smoothly, but rather shows a plateau-like structure between 538 and 540 eV.
The picture changes along the (001) plane. 
In this case, the spectra from O1 and O2 are quite similar to each other, except for a slight energy shift.
On the other hand, in the ELNES from O3 the two main peaks at $\sim$535 and $\sim$542 eV are considerably smoothed out and overall the oscillator strength is significantly lower compared to the other contributions.
These results confirm the sensitivity of ELNES to the local environment of the inequivalent oxygen atoms, enhanced in specific diffraction conditions. 
As a consequence of the anisotropy of the system, the weight of the transitions contributing to the spectra is significantly affected by the crystal orientation.

%%%%%%%%%%%%%%%%%%%%%%%%%%%%%%%%%%%%%%%%%%%%%%%%%%%%%
%  CONCLUSIONS
%%%%%%%%%%%%%%%%%%%%%%%%%%%%%%%%%%%%%%%%%%%%%%%%%%%%%
\section{Summary and Conclusions}
We have presented a joint theoretical and experimental study of core-level excitations from O 1$s$ electrons in $\beta$-\ce{Ga2O3}.
With a detailed \textit{ab initio} analysis of the electronic structure, we have characterized the conduction region in terms of band character, showing significant hybridization effects between O-$p$ and Ga-$s$ and -$p$ states. 
These features are reflected in the core-excitation absorption spectrum. 
Excitonic effects redistribute the oscillator strength to lower energy, and red-shift the first intense peak by approximately 2 eV. 
Moreover, bound excitons are formed at the spectral onset, the first one having a binding energy of about 0.5 eV.
Inequivalent O atoms give rise to different contributions in terms of excitation energies and intensity.
With this knowledge, we have analyzed the corresponding ELNES spectra, measured and computed with respect to different crystal planes.
The overall agreement between theory and experiment confirms the importance of an explicit treatment of many-body effects in $\beta$-\ce{Ga2O3}, and, in general, in TCOs. 
From the individual atomic contributions we have been able to identify the ELNES fingerprints of inequivalent oxygen atoms by considering specific crystallographic directions.

We can thus conclude that ELNES signals from atoms in specific environments can be selectively enhanced or quenched by appropriate adjustment of crystal orientation and diffraction conditions.
\textit{Ab initio} many-body theory, which allows for a reliable prediction and interpretation of these features, complements ELNES and ELCE  in view of characterizing anisotropic crystals with sensitivity to the atomic environment.
These results pave the way for further investigations of the local atomic environment of complex materials. 
In particular STEM-EELS measurements, which allow for probing the electronic structure at inequivalent atomic columns, are expected to provide additional insight.
TCOs are the ideal platform to exploit the potential of the proposed approach, which can substantially contribute to clarify the interplay between their crystal geometry and electronic structure.
In general, we consider our approach highly promising to characterize complex materials, including defected structures, alloys, interfaces, and heterostructures, in view of revealing signatures of non-equivalent atomic species with different chemical coordination. 

%%%%%%%%%%%%%%%%%%%%%%%%%%%%%%%%%%%%%%%%%%%%%%%%%%%%%
%  ACKNOWLEDGMENT
%%%%%%%%%%%%%%%%%%%%%%%%%%%%%%%%%%%%%%%%%%%%%%%%%%%%%
\section*{Acknowledgement}
This work was partially funded by the German Research Foundation (DFG), through the collaborative research center SFB 951. C.C.~acknowledges financial support from the \textit{Berliner Chancengleichheitsprogramm} (BCP). This work is embedded in the Leibniz Science Campus GraFOx (Growth and fundamentals of oxides for electronic applications).
%%%%%%%%%%%%%%%%%%%%%%%%%%%%%%%%%%%%%%%%%%%%%%%%%%%%%

%%%%%%%%%%%%%%%%%%%%%%%%%%%%%%%%%%%%%%%%%%%%%%%%%%%%%
\end{document}